%% file: draft.tex
\RequirePackage{lineno}
\documentclass[twocolumn,showpacs,superscriptaddress,amsmath,amssymb]{revtex4-1}
\usepackage[colorlinks,linkcolor=blue,anchorcolor=blue,citecolor=blue,urlcolor=blue]{hyperref}
\usepackage{graphicx}
\usepackage{epsfig}
\usepackage{dcolumn}
\usepackage{bm}
\usepackage{overpic}
\usepackage{color}
\usepackage{multirow}
\usepackage{subfigure}

\include{predef}

\begin{document}
\title{\boldmath Measurement of the Cross Section for $\ee\to$~Hadrons \\at Energies from 2.2324 to 3.6710 GeV}
\input{authors}

\date{\today}

\begin{abstract}
Based on electron-positron collision data collected with the BESIII detector operating at the Beijing Electron-Positron Collider II storage rings, the value of $R\equiv\sigma(\ee\to\textmd{hadrons})/\sigma(\ee\to\mm)$ is measured at 14 center-of-mass energies from 2.2324 to 3.6710 GeV. The resulting uncertainties are less than $3.0\%$, and are dominated by systematic uncertainties.
\end{abstract}

\maketitle

The lowest-order cross section of $\ee\to$ hadrons, known as the $R$ value in units of the cross section of the QED process $\ee\to\mm$, is a very important quantity in particle physics~\cite{bes3book}. Precision measurements of the $R$ value below $5~\gev$ contribute to the standard model (SM) prediction of the muon anomalous magnetic moment~\cite{review2020}. The latest direct measurement from Fermilab, when combined with previous measurements, increases the discrepancy between experiment and SM theory to 4.2 standard deviations~\cite{gminus2exp}. The discrepancy is reduced when experiment is compared with recent lattice gauge theory predictions~\cite{review2020,latticecal}. The $R$ value also contributes in the determination of the QED running coupling constant evaluated at the $Z$ pole, i.e., $\alfmz$. This observable provides another precision test for the SM and is essential for the electroweak precision physics program at future colliders. Many experiments have measured the $R$ value at low energies~\cite{Rvalue1,RvalueGamma2,Rvalue3,RvalueMarkI,RvaluePluto,slac4160,R2000,R2002,R2004,R2006A,R2006B,R_KEDR,R_KEDR2,R_KEDR3}, and the latest measurement with the KEDR detector has achieved a precision of better than 3.0\% above $3.1~\gev$~\cite{R_KEDR3}.

In this Letter, the $R$ value is measured at 14 center-of-mass (c.m.) energies ($\sqs$) from 2.2324 to $3.6710~\gev$, where the datasets were collected with the BESIII detector~\cite{BESIII} at Beijing Electron-Positron Collider II. The detector has a geometrical acceptance of 93\% of the 4$\pi$ solid angle. It contains a small-celled, helium-based main drift chamber (MDC), a time-of-flight system (TOF) based on plastic scintillators, an electromagnetic calorimeter (EMC) made of CsI(Tl) crystals, a muon system made of resistive plate chambers, and a superconducting solenoid magnet.

Experimentally, the $R$ value is determined with
\begin{equation}
\label{Rformula}
R=\frac{\Nhadobs-\Nbkg}{\csdimuborn\lint\efftrg\effhad(1+\delta)},
\end{equation}
where $\Nhadobs$ is the number of hadronic events directly counted from data, $\Nbkg$ is the number of background events remaining after all selection requirements, $\csdimuborn=4\pi\alpha^{2}(0)/(3s)$ is the leading-order cross section of $\ee\to\mm$, $\lint$ is the integrated luminosity of the data sample, $\efftrg$ is the trigger efficiency for hadronic events, $\effhad$ is the detection efficiency for inclusive hadronic events, and ($1+\delta$) is the initial state radiation (ISR) correction factor.

As a first step, $\ee\to\ee$ or $\ee\to\gaga$ events are identified and then rejected by requiring (i) at least two showers in the EMC, (ii) the angular difference $|\Delta\theta|=|\theta_{1}+\theta_{2}-180^{\circ}|$ less than $10^{\circ}$, where $\theta_{1,2}$ are the polar angles of the two most energetic showers, and (iii) the energy deposition of the second-most energetic shower of the event larger than $0.65\ebm$, where $\ebm$ represents the mean energy of the colliding beams.

For the remaining events, good charged hadronic tracks, which are referred to as "prongs" hereafter, are selected by imposing the following requirements: (i) the distance of the closest approach of these tracks to the interaction point ($\Vz$, $\Vr$) is required to be within 5 cm along and 0.5 cm perpendicular to the symmetry axis of the MDC; (ii) the charged tracks must lie in the acceptance region of the MDC, i.e. $|\!\cos\theta|<0.93$, where $\theta$ is the polar angle of the tracks; (iii) charged tracks that do not originate from the electron-positron collision, such as deuterons, are removed by $\chiP<10$ with $\chiP=(\dedx-\dedx_{p})/\sigma_{p}$, where $\dedx$ is the average energy loss directly measured in the MDC and $\dedx_{p}$ is the corresponding value expected for protons with a predicted uncertainty $\sigma_{p}$; (iv) the momenta of these charged tracks should be smaller than $0.94\pbm$, where $\pbm$ denotes the mean momentum of the colliding beams; and (v) charged tracks with both $E/(pc)>0.8$ and $p>0.65\pbm$ are removed, where $E$ is the deposited energy in the EMC and $p$ is the momentum measured by the MDC. Furthermore, any two oppositely charged tracks both with $E/(pc)>0.8$ will be regarded as arising from the gamma-conversion process~\cite{gamcon} and excluded as long as their invariant mass is less than $0.1~\gevcc$ and their opening angle is less than $15^{\circ}$.

To select hadronic events, at least two good charged tracks are required. For two-prong events, the two charged tracks should not be back-to-back, and the corresponding number of isolated photons ($N_{\textmd{iso}}^{\textmd{2-prg}}$) should be larger than 1. Two charged tracks are regarded to be back-to-back as long as both $|\Delta\theta|=|\theta_{1}+\theta_{2}-180^{\circ}|<10^{\circ}$ and $|\Delta\phi|=||\phi_{1}-\phi_{2}|-180^{\circ}|<15^{\circ}$ are satisfied, where $\theta$ and $\phi$ are the polar and azimuthal angles of the charged tracks, respectively. An isolated photon is selected from showers in the EMC by requiring the deposited energy to be larger than $100~\mev$, the opening angle to any charged track to exceed $20^{\circ}$, and the timing of the shower within 700 ns from the start time of the event. For three-prong events, the two charged tracks with highest and second highest momentum must not be back-to-back, and the number of charged tracks with $E/(pc)>0.8$ should be less than 2. In addition, the number of charged tracks with $\rpid>0.25$ must be less than 2, where $\rpid=\mathcal{P}(e)/\sum_{i}\mathcal{P}(i)$, with $i=e,~\pi,~K,~p$ and where $\mathcal{P}(i)$ is the particle identification (PID) probability calculated by combing MDC and TOF information. Events with more than three prongs are directly counted as hadronic events without any additional requirement.

Although the above inclusive hadronic selection criteria are applied, there are still residual background events contributing to $\Nhadobs$, coming from lepton pair production, two-photon processes, and beam-associated processes. Simulated data samples produced with a {\sc geant4}-based~\cite{geant4} Monte Carlo (MC) package, which includes the geometric description of the BESIII detector and the detector response, are used to estimate the background yields. The $\ee\to\ee$, $\gaga$, and $\mm$ processes are generated by {\sc babayaga3.5}~\cite{babayagav3.5}, while the $\ee\to\tata$ process is simulated by {\sc kkmc}~\cite{KKMC} with the subsequent decays of $\tau$ modeled by {\sc evtgen}~\cite{EVTGEN}. The two-photon processes $\ee\to\ee X$ with $X=\ee, \mm, \eta,~\eta^{\prime}, \pipi$, and $K^{+}K^{-}$~\cite{two-photon} are simulated using the generators {\sc diag36}~\cite{diag36}, {\sc ekhara}~\cite{ekhara} and {\sc galuga2.0}~\cite{galuga}.

Beam-associated background events, which originate from beam-gas interactions and Touschek scatterings, could be misidentified as signal events~\cite{BESIII}. An event vertex along the beam direction $\evz$, defined as the average of $\Vz$ of all the good charged tracks, and another one $\evzloose$, taking into account also those tracks not satisfying the $\Vz$ requirement, are used to estimate the number of the beam-associated events. In the $|\evzloose|$ distribution, (0, 5)~cm is taken as the signal region and (5, 10)~cm as the sideband. All events in the sideband region are assumed to be beam associated. Alternatively, the $\evz$ distribution is fitted with a double Gaussian as signal plus a background shape that is obtained from separated-beam data. The two methods give consistent results, which are also confirmed at $\sqs=2.4000$ and $3.4000~\gev$ by directly analyzing the separated-beam data and scaling
according to the corresponding data-taking time.

The integrated luminosity is determined by analyzing large-angle Bhabha events at each c.m.~energy~\cite{luminosity}. The trigger efficiency for hadronic events is nearly $100\%$~\cite{trigger}.

\begin{figure}[!htbp]
\setlength{\abovecaptionskip}{0.2cm}
\setlength{\belowcaptionskip}{-0.5cm}
\begin{center}
\begin{overpic}[width=1.68in]{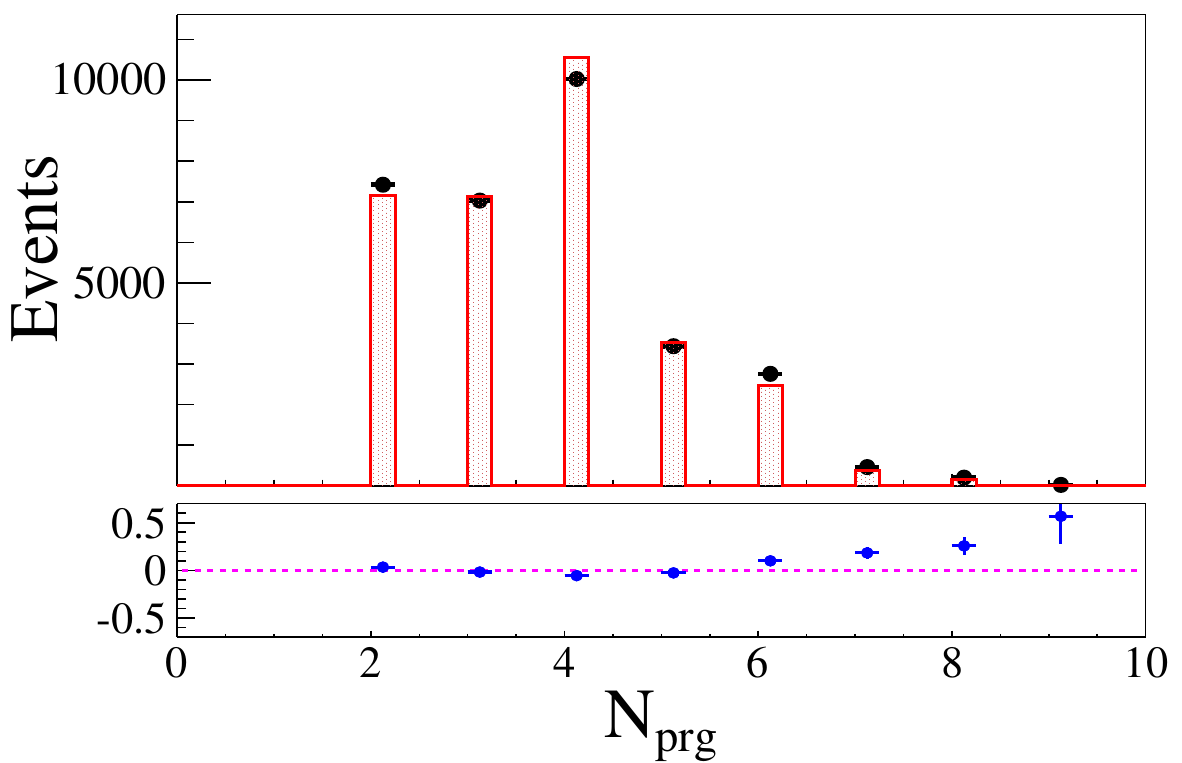}
\end{overpic}
\begin{overpic}[width=1.68in]{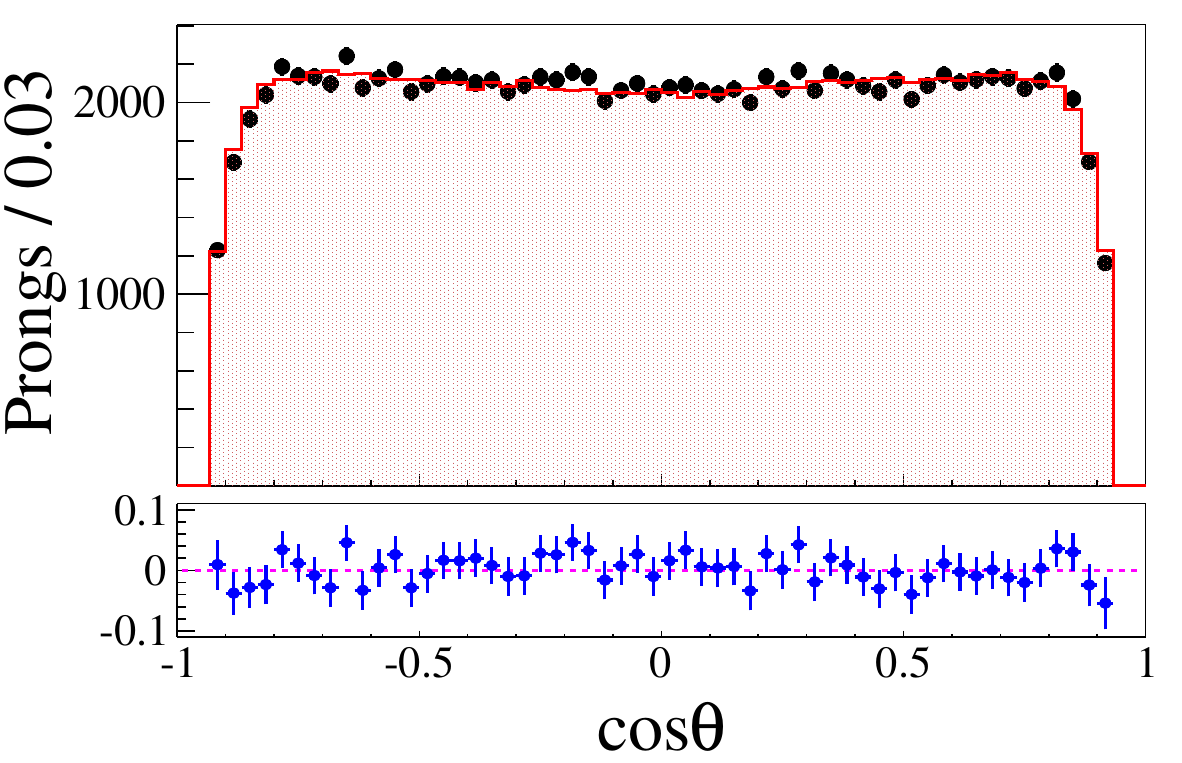}
\end{overpic}
\begin{overpic}[width=1.68in]{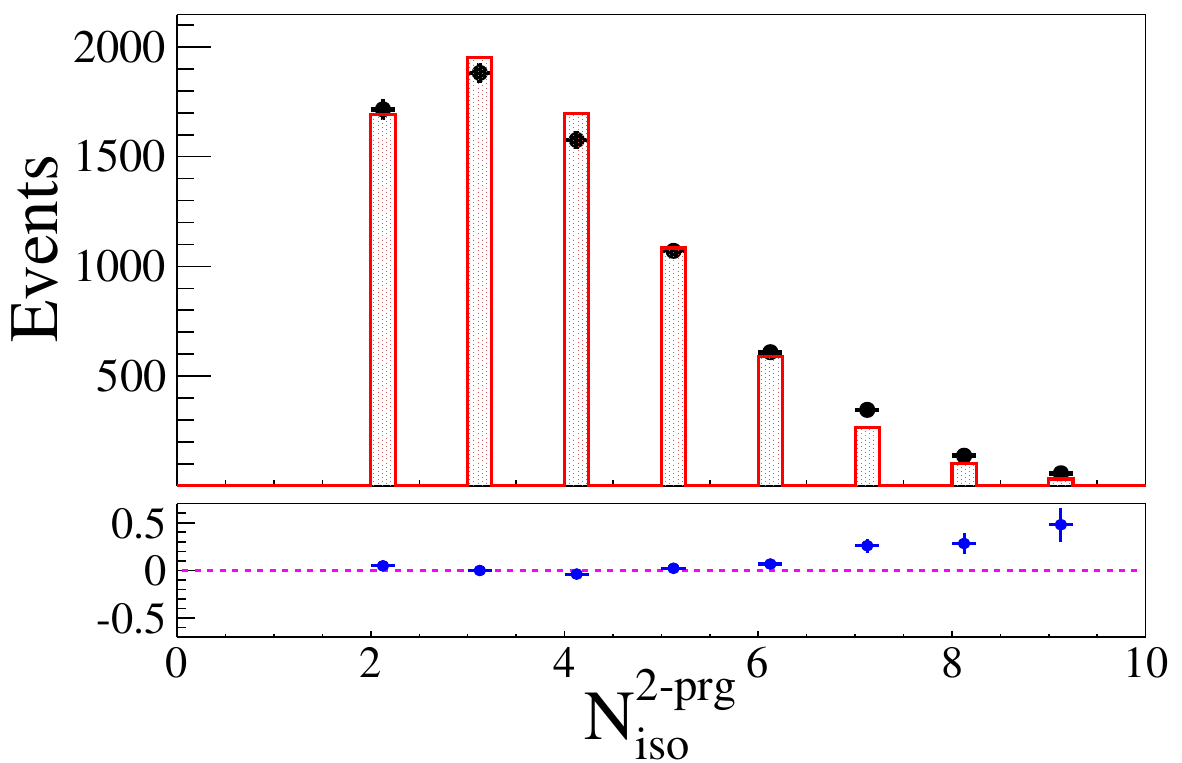}
\end{overpic}
\begin{overpic}[width=1.68in]{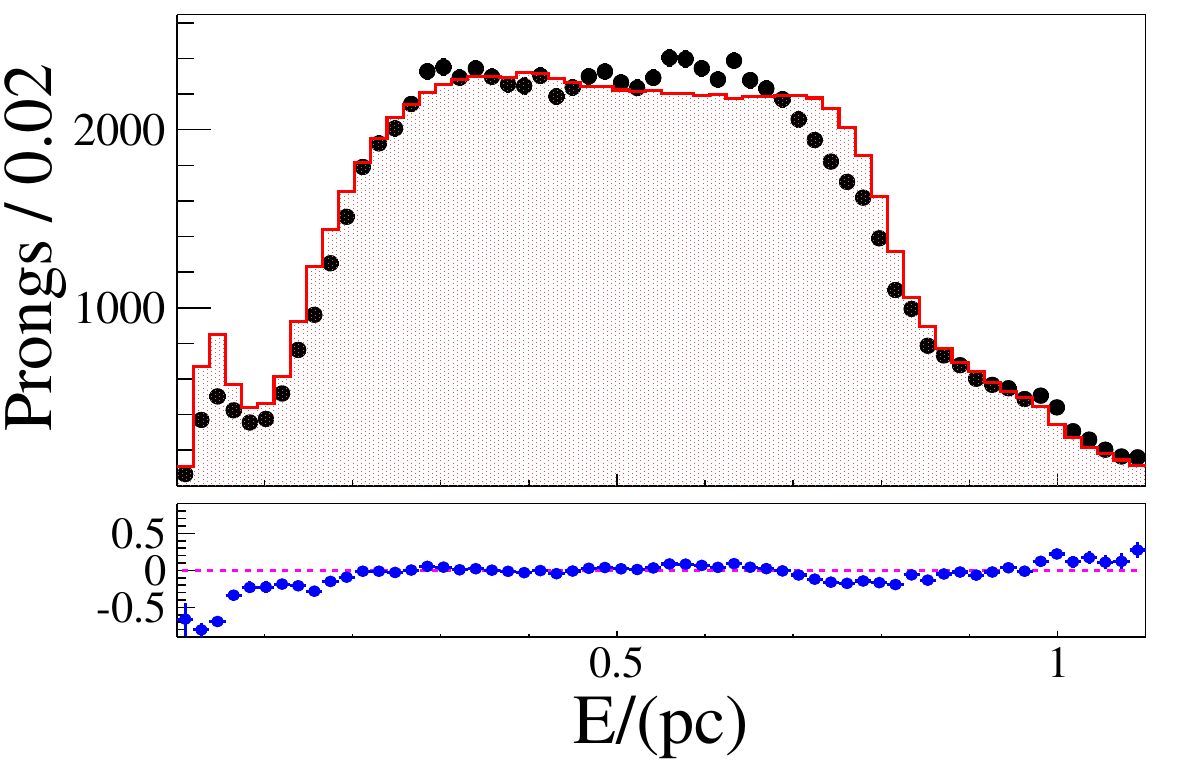}
\end{overpic}
\caption{Comparison between data (black dots) and a simulation based on the {\sc luarlw} MC generator (red histograms) at $\sqs=3.4000~\gev$. Here, $N_{\textmd{prg}}$ is the number of prongs, and $N_{\textmd{iso}}^{\textmd{2-prg}}$ is the number of isolated photons in two-prong events. Track-level variables $\theta$, $E$, and $p$ stand for the polar angle, deposited energy in EMC, and MDC measured momentum of each prong. The MC simulation distribution is normalized to that in data by the ratio of the corresponding numbers of total events or tracks. The blue dots denote the relative differences.
}
\label{DisAt34000}
\end{center}
\end{figure}

To estimate the hadronic detection efficiency, the {\sc luarlw} model, developed to simulate inclusive hadronic events~\cite{lundmodel, FewBody}, is used. After the ISR process, resonance and continuum processes are produced according to the respective cross sections at the effective c.m.~energy ($\sqsp$). All allowed $1^{--}$ resonances, including $\rho(770)$, $\omega(782)$, $\phi(1020)$, $J/\psi$, and their excited states are implemented in the {\sc luarlw} generator with corresponding resonance parameters taken from the Particle Data Group (PDG)~\cite{pdg2020}. In the {\sc luarlw} generator, the conventional resonances decay according to the branching fractions listed by the PDG, while the hadronic as well as some radiative decays of $J/\psi$ meson and the production of continuum process at any $\sqsp$ are modeled by the Lund area law~\cite{FewBody}. In this area law, a group of parameters controlling the sampling of flavors, multiplicities, and kinematic quantities of generated initial hadrons are tuned according to extensive comparisons between simulation and data in a large number of variables. The agreement after tuning is demonstrated in Fig.~\ref{DisAt34000}, where the signal MC sample is compared with data in some of the critical distributions at $\sqs=3.4000~\gev$ after subtracting the QED-related and beam-associated background contributions. Further comparisons show that the tuned {\sc luarlw} MC generator can reasonably reproduce inclusive hadronic events in experimental data at all the c.m.~energies utilized in this Letter.

The Feynman diagram (FD) scheme is used to simulate the ISR process in the {\sc luarlw} generator and calculate the correction factors $(1+\delta)$ after formalizing the contribution of the vacuum polarization (VP) correction with the complex expression $1/|1-\Pi(s)|^{2}$ instead of the four real terms used in Ref.~\cite{slac4160},
\begin{equation}
\label{fdnominalisr}
1+\delta=\frac{\delvert}{|1-\Pi(s)|^{2}}+\int_{0}^{x_{m}}\frac{F(x,s)}{|1-\Pi(s^{\prime})|^{2}}\frac{\cshadborn(s^{\prime})}{\cshadborn(s)}dx .
\end{equation}
Here, $x=1-s^{\prime}/s$ and $x_{m}=1-4m_{\pi}^{2}/s$ is the maximum allowed energy fraction of the ISR photon in the process $\ee\to\pipi$, $\delvert$ accounts for the contribution of the initial state vertex correction, and $F(x,s)$ denotes the radiator function of the FD scheme~\cite{slac4160}. Contributions of leptons and narrow resonances to the VP operator $\Pi(s)$ are calculated analytically~\cite{greiner, shamov}, while that of the remaining hadronic processes is determined via the dispersion relation~\cite{slac4160} and the optical theorem~\cite{R_KEDR}. Since $\cshadborn(s)$, which parametrizes the lowest-order hadronic cross section from both the continuum and resonance processes, is used as input to $\Pi(s)$ and $(1+\delta)$, an iterative procedure is performed based on the relation $\cshadborn=R\csdimuborn$. The final $R$ value is obtained when the iteration converges.

\linespread{1.1}
\begin{table*}[!htbp]
\setlength{\abovecaptionskip}{3pt}
\setlength{\belowcaptionskip}{-5pt}
\caption{Summary of systematic uncertainties (in percent) at each c.m.~energy, where the total uncertainty is the sum of the individual ones in quadrature. Uncertainties from the last four sources are correlated between the energy points.}
\label{rvaluesys}
\begin{center}
\setlength{\tabcolsep}{3.2mm}{
\begin{tabular}{ c  c  c  c  c  c  c  c  c }
\hline
\hline
\multicolumn{1}{  c  }{\multirow{1}{*}{} } &
\multicolumn{1}{  c  }{\multirow{1}{*}{Event} } &
\multicolumn{1}{  c  }{\multirow{1}{*}{QED} } &
\multicolumn{1}{  c  }{\multirow{1}{*}{Beam} } &
\multicolumn{1}{  c  }{\multirow{1}{*}{} } &
\multicolumn{1}{  c  }{\multirow{1}{*}{Trigger} } &
\multicolumn{1}{  c  }{\multirow{1}{*}{Signal} } &
\multicolumn{1}{  c  }{\multirow{1}{*}{ISR} } &
\multicolumn{1}{  c  }{\multirow{1}{*}{} } \\
\multicolumn{1}{  c  }{\multirow{1}{*}{$\sqs$~($\gev$)} } &
\multicolumn{1}{  c  }{\multirow{1}{*}{selection} } &
\multicolumn{1}{  c  }{\multirow{1}{*}{background} } &
\multicolumn{1}{  c  }{\multirow{1}{*}{background} } &
\multicolumn{1}{  c  }{\multirow{1}{*}{Luminosity} } &
\multicolumn{1}{  c  }{\multirow{1}{*}{efficiency} } &
\multicolumn{1}{  c  }{\multirow{1}{*}{model} } &
\multicolumn{1}{  c  }{\multirow{1}{*}{correction} } &
\multicolumn{1}{  c  }{\multirow{1}{*}{Total} } \\ \cline{1-9}
\multicolumn{1}{  c  }{ 2.2324 } &  0.41  &  0.23  &  0.28  &  0.80  &  0.10  &  0.60  &  1.15  &  1.62  \\ 
\multicolumn{1}{  c  }{ 2.4000 } &  0.55  &  0.27  &  0.15  &  0.80  &  0.10  &  1.11  &  1.10  &  1.87  \\ 
\multicolumn{1}{  c  }{ 2.8000 } &  0.58  &  0.28  &  0.34  &  0.80  &  0.10  &  1.97  &  1.06  &  2.48  \\ 
\multicolumn{1}{  c  }{ 3.0500 } &  0.61  &  0.33  &  0.41  &  0.80  &  0.10  &  1.76  &  1.01  &  2.33  \\ 
\multicolumn{1}{  c  }{ 3.0600 } &  0.60  &  0.34  &  0.48  &  0.80  &  0.10  &  1.84  &  1.00  &  2.39  \\ 
\multicolumn{1}{  c  }{ 3.0800 } &  0.61  &  0.35  &  0.35  &  0.80  &  0.10  &  1.31  &  1.05  &  2.02  \\ 
\multicolumn{1}{  c  }{ 3.4000 } &  0.65  &  0.33  &  0.16  &  0.80  &  0.10  &  1.86  &  1.24  &  2.49  \\ 
\multicolumn{1}{  c  }{ 3.5000 } &  0.60  &  0.35  &  0.62  &  0.80  &  0.10  &  2.05  &  1.16  &  2.66  \\ 
\multicolumn{1}{  c  }{ 3.5424 } &  0.61  &  0.37  &  0.01  &  0.80  &  0.10  &  2.05  &  1.14  &  2.58  \\ 
\multicolumn{1}{  c  }{ 3.5538 } &  0.66  &  0.31  &  0.39  &  0.80  &  0.10  &  2.22  &  1.13  &  2.74  \\ 
\multicolumn{1}{  c  }{ 3.5611 } &  0.74  &  0.34  &  0.34  &  0.80  &  0.10  &  2.28  &  1.12  &  2.81  \\ 
\multicolumn{1}{  c  }{ 3.6002 } &  0.66  &  0.33  &  0.38  &  0.80  &  0.10  &  2.27  &  1.09  &  2.77  \\ 
\multicolumn{1}{  c  }{ 3.6500 } &  0.53  &  0.35  &  0.69  &  0.80  &  0.10  &  2.28  &  1.13  &  2.83  \\ 
\multicolumn{1}{  c  }{ 3.6710 } &  0.61  &  0.42  &  0.63  &  0.80  &  0.10  &  2.23  &  1.04  &  2.77  \\ 
\hline
\hline
\end{tabular}}
\end{center}
\end{table*}

The systematic uncertainties on the $R$ values are addressed according to the inputs in Eq.~(\ref{Rformula}). All selection criteria are varied around their nominal values to estimate corresponding changes in $\Nhadobs$, $\Nbkg$, and $\effhad$. The resulting deviations of the $R$ values, which are found to be less than 0.8\% for all c.m.~energies, are regarded as the systematic uncertainties due to the imperfect reproduction of the data by MC samples. For the uncertainties of the QED-related background estimation, cross checks are performed by using the \textsc{babayaga nlo}~\cite{babayaganlo} package for $\ee\to\ee$ as well as $\gaga$ processes, and the {\sc phokhara}~\cite{phokhara} generator for $\mm$ events. For the $\ee\to\tata$ process, the uncertainty of the {\sc kkmc} program is negligible, due to the accuracy of the simulation and the small fraction of this background. In addition, the missing two-photon channels like $\ee\to\ee\pi^{0}$ and the observed contributions from some intermediate states such as the $\rho(770)$ and $f_{2}(1270)$ in process $\ee\to\ee\pipi$, which are not included in the current two-photon MC samples, are also considered as systematic uncertainties. The total uncertainty of the $R$ values due to QED-related background processes is less than 0.5\% due to the small fraction of QED-related events in $\Nhadobs$. The difference in the number of the beam-associated background events estimated by the sideband and fit methods, which is found to be less than 0.7\%, is taken as the systematic uncertainty. The uncertainty on $\csdimuborn$ can be neglected since it can be exactly calculated in QED~\cite{greiner}. The uncertainties of the integrated luminosity and the trigger efficiency are obtained to be 0.8\%~\cite{luminosity} and 0.1\%~\cite{trigger}, respectively.

The most crucial sources of systematic uncertainties in this Letter originate from the simulation of the inclusive hadronic events and the calculation of the ISR correction factors. To estimate the corresponding uncertainties, a composite generator, referred to here as "{\sc hybrid}", has been developed and investigated intensively~\cite{hybrid}. The {\sc hybrid} generator integrates the {\sc conexc}~\cite{conexc}, {\sc phokhara}, and {\sc luarlw} models. {\sc conexc} simulates a total of 47 exclusive processes according to a homogeneous and isotropic phase-space population, but reproducing the measured line shapes of the absolute cross section. {\sc phokhara} generates ten well-parametrized and established exclusive channels, including $\ee\to\pipi$, $\pipi\pi^{0}$, $\pipi\pi^{0}\pi^{0}$, and $\pipi\pipi$~\cite{phokhara}. The remaining unknown decays of the virtual photon are modeled by {\sc luarlw}. To implement as much experimental knowledge as currently available, processes containing intermediate states such as $K^{\ast}(892)$, $\omega(782)$, $\phi(1020)$, and $\eta$ are implemented in the {\sc conexc} component after subtracting their contributions to related inclusive processes to avoid double counting~\cite{egconexc}. In addition, a veto procedure has been implemented in the {\sc hybrid} generator to avoid {\sc luarlw} reproducing exclusive processes that are already generated with {\sc conexc} or {\sc phokhara}. Finally, a set of chosen parameters in the {\sc luarlw} component are iteratively tuned and a good consistency between the {\sc hybrid} simulation and data is achieved. The resulting deviations in $\effhad$ between the nominal {\sc luarlw} generator and the {\sc hybrid} generator are less than 1.3\%. In {\sc hybrid}, the structure function scheme~\cite{hybrid,nicrosini1989} is used to simulate the ISR process and calculate the corresponding correction factors, where the vacuum polarization operator $\Pi(s)$ is parametrized in Ref.~\cite{jegerlehnervp}. The maximum difference of the calculated $(1+\delta)$ factor between the {\sc hybrid} and {\sc luarlw} simulations is 1.4\%, which is mainly due to the different parametrization schemes of the ISR process. Since the hadronic detection efficiency is correlated with the ISR correction, the deviations of the resulting $R$ values between these two simulation schemes, which are less than 2.3\% after implementing the same $\cshadborn$ line shape, are regarded as systematic uncertainties. In addition to the different calculation schemes, the accuracy of the FD scheme and the uncertainty of the $\cshadborn$ line shape are considered as uncertainties of the ISR correction factors; the maximum value is less than 1.3\% for all the c.m.~energies. The systematic uncertainties at each c.m.~energy are summarized in Table~\ref{rvaluesys}.

Table~\ref{rvaluetab} lists the primary quantities mentioned in Eq.~(\ref{Rformula}) for each energy; the total uncertainty is less than 3.0\% and is dominated by systematic effects.

\linespread{1.0}
\begin{table*}[!htbp]
\setlength{\abovecaptionskip}{3pt}
\setlength{\belowcaptionskip}{-5pt}
\caption{Summary of primary quantities mentioned in Eq.~(\ref{Rformula}) and the measured $R$ value for each c.m.~energy, where the uncertainties are statistical and systematic respectively.}
\label{rvaluetab}
\begin{center}
\setlength{\tabcolsep}{4mm}{
\begin{tabular}{ c | r  r  r  c  c  c  c}
\hline
\hline
\multicolumn{1}{  c  }{\multirow{1}{*}{$\sqs$~($\gev$)} } &
\multicolumn{1}{  c  }{\multirow{1}{*}{$\Nhadobs$} } &
\multicolumn{1}{  c  }{\multirow{1}{*}{$\Nbkg$} } &
\multicolumn{1}{  c  }{\multirow{1}{*}{$\csdimuborn$~(nb)} } &
\multicolumn{1}{  c  }{\multirow{1}{*}{$\lint$~$(\textmd{pb}^{-1})$} } &
\multicolumn{1}{  c  }{\multirow{1}{*}{$\effhad$~(\%)} } &
\multicolumn{1}{  c  }{\multirow{1}{*}{$1+\delta$} } &
\multicolumn{1}{  c  }{\multirow{1}{*}{$R$} } \\ \cline{1-8}
\multicolumn{1}{  c  }{ 2.2324 } &  83227~ & 2041~ & 17.427~~ & 2.645 & 64.45 & 1.195 & $2.286\pm0.008\pm0.037$ \\ 
\multicolumn{1}{  c  }{ 2.4000 } &  96627~ & 2331~ & 15.079~~ & 3.415 & 67.29 & 1.204 & $2.260\pm0.008\pm0.042$ \\ 
\multicolumn{1}{  c  }{ 2.8000 } &  83802~ & 2075~ & 11.078~~ & 3.753 & 72.25 & 1.219 & $2.233\pm0.008\pm0.055$ \\ 
\multicolumn{1}{  c  }{ 3.0500 } &  283822~ & 7719~ & 9.337~~ & 14.89~~~ & 73.91 & 1.193 & $2.252\pm0.004\pm0.052$ \\ 
\multicolumn{1}{  c  }{ 3.0600 } &  282467~ & 7683~ & 9.276~~ & 15.04~~~ & 73.88 & 1.183 & $2.255\pm0.004\pm0.054$ \\ 
\multicolumn{1}{  c  }{ 3.0800 } &  552435~ & 15433~ & 9.156~~ & 31.02~~~ & 73.98 & 1.123 & $2.277\pm0.003\pm0.046$ \\ 
\multicolumn{1}{  c  }{ 3.4000 } &  32202~ & 843~ & 7.513~~ & 1.733 & 74.81 & 1.382 & $2.330\pm0.014\pm0.058$ \\ 
\multicolumn{1}{  c  }{ 3.5000 } &  62670~ & 1691~ & 7.090~~ & 3.633 & 75.32 & 1.351 & $2.327\pm0.010\pm0.062$ \\ 
\multicolumn{1}{  c  }{ 3.5424 } &  145303~ & 3872~ & 6.921~~ & 8.693 & 75.58 & 1.341 & $2.319\pm0.006\pm0.060$ \\ 
\multicolumn{1}{  c  }{ 3.5538 } &  92996~ & 2469~ & 6.877~~ & 5.562 & 75.50 & 1.338 & $2.342\pm0.008\pm0.064$ \\ 
\multicolumn{1}{  c  }{ 3.5611 } &  64650~ & 2477~ & 6.849~~ & 3.847 & 75.50 & 1.337 & $2.338\pm0.010\pm0.066$ \\ 
\multicolumn{1}{  c  }{ 3.6002 } &  159644~ & 9817~ & 6.701~~ & 9.502 & 75.73 & 1.328 & $2.339\pm0.006\pm0.065$ \\ 
\multicolumn{1}{  c  }{ 3.6500 } &  78730~ & 6168~ & 6.519~~ & 4.760 & 76.00 & 1.308 & $2.352\pm0.009\pm0.067$ \\ 
\multicolumn{1}{  c  }{ 3.6710 } &  75253~ & 6461~ & 6.445~~ & 4.628 & 76.11 & 1.260 & $2.405\pm0.010\pm0.067$ \\ 
\hline
\hline
\end{tabular}}
\end{center}
\end{table*}

Some additional efforts are made to check the reliability of the $R$ values obtained in this Letter. Dedicated selection criteria are developed to include one-prong events, and the resulting deviation from the nominal $R$ value is at most 0.8\%. On the other hand, the exclusion of two-prong events changes the $R$ value by a maximum of 2.2\%. To quantitatively verify the {\sc luarlw} generator, some exclusive processes are investigated. Production fractions and efficiencies with the nominal inclusive hadronic event selection of the processes $\ee\to\pipi$ and $\pipi\pi^{0}$ are compared between that of {\sc luarlw} and {\sc hybrid} simulations, respectively. A hypothetical model is constructed by replacing the production fraction and inclusive efficiency of the exclusive process of interest in {\sc luarlw} with that of {\sc hybrid}, which gives a better reproduction of this process due to the inclusion of corresponding experimental knowledge. Differences of resulting $\effhad$ from the hypothetical and the nominal {\sc luarlw} models are less than 2.1\% and 0.6\% for $\pipi$ and $\pipi\pi^{0}$ processes, respectively. Similarly, the processes $\ee\to2(\pipi)$ and $\pipi2\pi^{0}$ are also studied but with production fractions extracted directly from data. The maximum deviations of $\effhad$ using these corresponding hypothetical models are found to be 1.1\% and 0.6\%, respectively. Furthermore, $\effhad$ changes by 0.2\% at $\sqs=3.4000~\gev$ if the decays of $J/\psi$ mesons produced in {\sc luarlw} are modeled by the {\sc hybrid} generator, in which a comparably accurate description of the data is observed. The ISR correction factors are also calculated by the structure function scheme mentioned in Ref.~\cite{kureavfadin}, and the maximum deviation to the nominally applied FD scheme is 1.3\%. The quantity $\effhad(0)(1+\delobs)$ used in a different $R$ value measurement method in Refs.~\cite{R2000,R2002,R2004} is also calculated, which differs from $\effhad(1+\delta)$ used in this Letter by 0.8\% at most.  The deviations observed in these checks are not taken as additional contributions to the systematic uncertainties since they are already covered by the previously discussed systematic uncertainties.

\begin{figure}[!htbp]
\setlength{\abovecaptionskip}{-0.0cm}
\setlength{\belowcaptionskip}{-0.4cm}
\begin{center}
\begin{overpic}[width=3.4in]{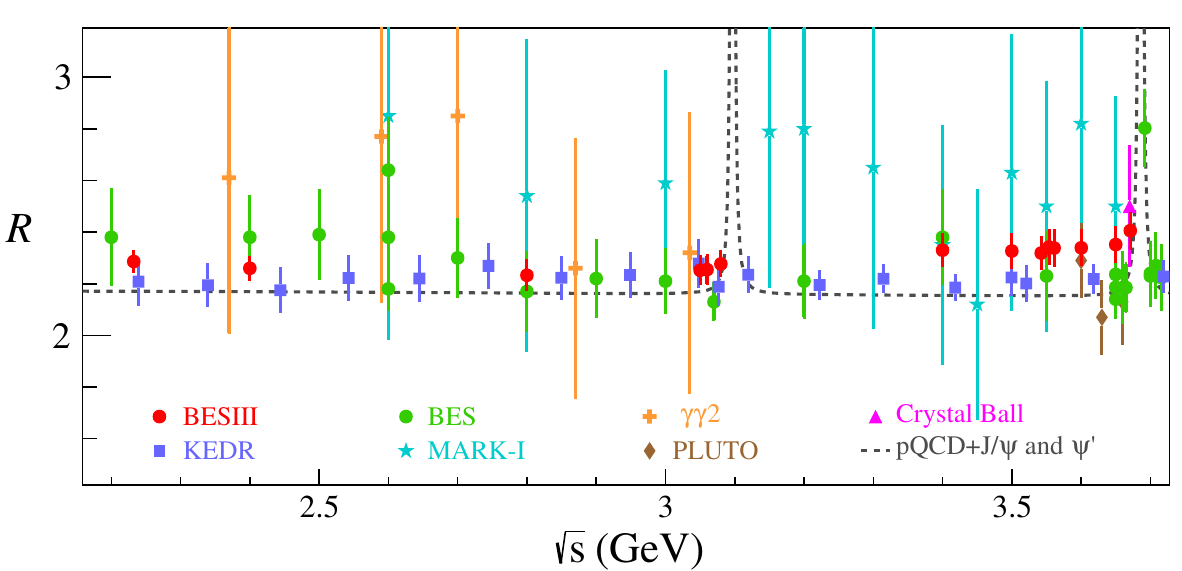}
\end{overpic}
\caption{Comparison of $R$ values in the c.m.~energy region from 2.2 to $3.7~\gev$, where the red dots denote that of BESIII, green dots stand for that of BES~\cite{R2000,R2002,R2004,R2006A,R2006B}, rectangles show KEDR measurements~\cite{R_KEDR,R_KEDR2,R_KEDR3}, orange crosses are $R$ values from the $\gamma\gamma2$ Collaboration~\cite{RvalueGamma2}, cyan stars are that of MARK-I~\cite{RvalueMarkI}, brown diamonds are PLUTO results~\cite{RvaluePluto}, and the $R$ value of the Crystal Ball Collaboration is shown as a magenta triangle~\cite{slac4160}.}
\label{rvalueplot}
\end{center}
\end{figure}

Figure~\ref{rvalueplot} shows the $R$ value obtained in this analysis, together with previous measurements~\cite{RvalueGamma2,RvalueMarkI,RvaluePluto,slac4160,R2000,R2002,R2004,R2006A,R2006B,R_KEDR,R_KEDR2,R_KEDR3}. A theoretical expectation of $R$ obtained by combining the perturbative QCD prediction~\cite{qcdsumrule} and the contributions from involved narrow resonances is also illustrated with the dashed curve in Fig.~\ref{rvalueplot}. The $R$ values from BESIII have an accuracy of better than 2.6\% below $3.1~\gev$ and 3.0\% above. The average $R$ value in the c.m.~energy range $3.4\sim3.6~\gev$ obtained by BESIII is larger than the corresponding KEDR result and theoretical expectation by 1.9 and 2.7 standard deviations (accounting for 100\% correlated systematics from the last four of the seven contributions in Table~\ref{rvaluesys}), respectively. Further precision measurements are desired and will help to improve the accuracy of the SM predictions of $\alfmz$, as well as the muon magnetic anomaly, and to verify the QCD sum rules at lower energies~\cite{qcdsumrule}.

The BESIII Collaboration thanks the staff of BEPCII, the IHEP computing center and the supercomputing center of USTC for their strong support. This work is supported in part by National Key R\&D Program of China under Contracts No. 2020YFA0406400, No. 2020YFA0406300; National Natural Science Foundation of China (NSFC) under Contracts No. 11335008, No. 11625523, No. 11635010, No. 11735014, No. 11822506, No. 11835012, No. 11935015, No. 11935016, No. 11935018, No. 11961141012, No. 12022510, No. 12025502, No. 12035009, No. 12035013, No. 12175244, No. 12061131003, No. 11705192, No. 11875115, No. 11875262, No. 11950410506, No. 12061131003; the Chinese Academy of Sciences (CAS) Large-Scale Scientific Facility Program; Joint Large-Scale Scientific Facility Funds of the NSFC and CAS under Contracts No. U1732263, No. U1832207, No. U1832103, No. U2032105, No. U2032111; CAS Key Research Program of Frontier Sciences under Contract No. QYZDJ-SSW-SLH040; 100 Talents Program of CAS; INPAC and Shanghai Key Laboratory for Particle Physics and Cosmology; ERC under Contract No. 758462; European Union Horizon 2020 Research and Innovation Programme under Marie Sklodowska-Curie Grant Agreement No. 894790; German Research Foundation DFG under Contracts No. 443159800, Collaborative Research Center CRC 1044, FOR 2359, FOR 2359, GRK 214; Istituto Nazionale di Fisica Nucleare, Italy; Ministry of Development of Turkey under Contract No. DPT2006K-120470; National Science and Technology fund; Olle Engkvist Foundation under Contract No. 200-0605; STFC (United Kingdom); The Knut and Alice Wallenberg Foundation (Sweden) under Contract No. 2016.0157; The Royal Society, UK under Contracts No. DH140054, No. DH160214; The Swedish Research Council; U. S. Department of Energy under Awards No. DE-FG02-05ER41374, No. DE-SC-0012069.

\input{Reference.tex}


\end{document}

%% file: predef.tex
\newcommand{\ee}{e^{+}e^{-}}
\newcommand{\mm}{\mu^{+}\mu^{-}}

\newcommand{\alfmz}{\alpha(M_{Z}^{2})}

\newcommand{\ebm}{E_{\textmd{beam}}}

\newcommand{\pbm}{p_{\textmd{beam}}}

\newcommand{\tata}{\tau^{+}\tau^{-}}
\newcommand{\pipi}{\pi^{+}\pi^{-}}
\newcommand{\gaga}{\gamma\gamma}

\newcommand{\sqs}{\sqrt{s}}
\newcommand{\sqsp}{\sqrt{s^{\prime}}}

\newcommand{\dedx}{\textmd{d}E/\textmd{d}x}

\newcommand{\chiP}{\chi_{p}}
\newcommand{\evz}{V_{z}^{\textmd{evt}}}
\newcommand{\evzloose}{V_{z,\textmd{loose}}^{\textmd{evt}}}

\newcommand{\rpid}{r_{\textmd{PID}}}

\newcommand{\Nhadobs}{N_{\textmd{had}}^{\textmd{obs}}}

\newcommand{\effhad}{\vap_{\textmd{had}}}
\newcommand{\efftrg}{\vap_{\textmd{trig}}}
\newcommand{\lint}{\mathcal{L}_{\textmd{int}}}
\newcommand{\Nbkg}{N_{\textmd{bkg}}}

\newcommand{\vap}{\varepsilon}

\newcommand{\cshadborn}{\sigma_{\textmd{had}}^{0}}

\newcommand{\csdimuborn}{\sigma_{\mu\mu}^{0}}

\newcommand{\delvert}{\delta_{\textmd{vert}}}

\newcommand{\delobs}{\delta_{\textmd{obs}}}

\newcommand{\Vr}{V_{r}}
\newcommand{\Vz}{V_{z}}

\newcommand{\gev}{\mathrm{GeV}}
\newcommand{\mev}{\mathrm{MeV}}

\newcommand{\gevcc}{\mathrm{GeV}/c^2}



%% file: authors.tex
\author{
\begin{small}
\begin{center}
M.~Ablikim$^{1}$, M.~N.~Achasov$^{10,b}$, P.~Adlarson$^{66}$, S. ~Ahmed$^{14}$, M.~Albrecht$^{4}$, R.~Aliberti$^{27}$, A.~Amoroso$^{65a,65c}$, M.~R.~An$^{31}$, Q.~An$^{62,48}$, X.~H.~Bai$^{56}$, Y.~Bai$^{47}$, O.~Bakina$^{28}$, R.~Baldini Ferroli$^{22a}$, I.~Balossino$^{23a}$, Y.~Ban$^{37,h}$, K.~Begzsuren$^{25}$, N.~Berger$^{27}$, M.~Bertani$^{22a}$, D.~Bettoni$^{23a}$, F.~Bianchi$^{65a,65c}$, J.~Bloms$^{59}$, A.~Bortone$^{65a,65c}$, I.~Boyko$^{28}$, R.~A.~Briere$^{5}$, H.~Cai$^{67}$, X.~Cai$^{1,48}$, A.~Calcaterra$^{22a}$, G.~F.~Cao$^{1,53}$, N.~Cao$^{1,53}$, S.~A.~Cetin$^{52a}$, J.~F.~Chang$^{1,48}$, W.~L.~Chang$^{1,53}$, G.~Chelkov$^{28,a}$, D.~Y.~Chen$^{6}$, G.~Chen$^{1}$, H.~S.~Chen$^{1,53}$, M.~L.~Chen$^{1,48}$, S.~J.~Chen$^{34}$, X.~R.~Chen$^{24}$, Y.~B.~Chen$^{1,48}$, Z.~J~Chen$^{19,i}$, W.~S.~Cheng$^{65c}$, G.~Cibinetto$^{23a}$, F.~Cossio$^{65c}$, X.~F.~Cui$^{35}$, H.~L.~Dai$^{1,48}$, X.~C.~Dai$^{1,53}$, A.~Dbeyssi$^{14}$, R.~ E.~de Boer$^{4}$, D.~Dedovich$^{28}$, Z.~Y.~Deng$^{1}$, A.~Denig$^{27}$, I.~Denysenko$^{28}$, M.~Destefanis$^{65a,65c}$, F.~De~Mori$^{65a,65c}$, Y.~Ding$^{32}$, C.~Dong$^{35}$, J.~Dong$^{1,48}$, L.~Y.~Dong$^{1,53}$, M.~Y.~Dong$^{1,48,53}$, X.~Dong$^{67}$, S.~X.~Du$^{70}$, Y.~L.~Fan$^{67}$, J.~Fang$^{1,48}$, S.~S.~Fang$^{1,53}$, Y.~Fang$^{1}$, R.~Farinelli$^{23a}$, L.~Fava$^{65B,65C}$, F.~Feldbauer$^{4}$, G.~Felici$^{22a}$, C.~Q.~Feng$^{62,48}$, J.~H.~Feng$^{49}$, M.~Fritsch$^{4}$, C.~D.~Fu$^{1}$, Y.~Gao$^{62,48}$, Y.~Gao$^{37,h}$, Y.~G.~Gao$^{6}$, I.~Garzia$^{23a,23b}$, P.~T.~Ge$^{67}$, C.~Geng$^{49}$, E.~M.~Gersabeck$^{57}$, A~Gilman$^{60}$, K.~Goetzen$^{11}$, L.~Gong$^{32}$, W.~X.~Gong$^{1,48}$, W.~Gradl$^{27}$, M.~Greco$^{65a,65c}$, L.~M.~Gu$^{34}$, M.~H.~Gu$^{1,48}$, C.~Y~Guan$^{1,53}$, A.~Q.~Guo$^{21}$, L.~B.~Guo$^{33}$, R.~P.~Guo$^{39}$, Y.~P.~Guo$^{9,f}$, A.~Guskov$^{28,a}$, T.~T.~Han$^{40}$, W.~Y.~Han$^{31}$, X.~Q.~Hao$^{15}$, F.~A.~Harris$^{55}$, K.~L.~He$^{1,53}$, F.~H.~Heinsius$^{4}$, C.~H.~Heinz$^{27}$, Y.~K.~Heng$^{1,48,53}$, C.~Herold$^{50}$, M.~Himmelreich$^{11,d}$, T.~Holtmann$^{4}$, G.~Y.~Hou$^{1,53}$, Y.~R.~Hou$^{53}$, Z.~L.~Hou$^{1}$, H.~M.~Hu$^{1,53}$, J.~F.~Hu$^{46,j}$, T.~Hu$^{1,48,53}$, Y.~Hu$^{1}$, G.~S.~Huang$^{62,48}$, L.~Q.~Huang$^{63}$, X.~T.~Huang$^{40}$, Y.~P.~Huang$^{1}$, Z.~Huang$^{37,h}$, T.~Hussain$^{64}$, N~H\"usken$^{21,27}$, W.~Ikegami Andersson$^{66}$, W.~Imoehl$^{21}$, M.~Irshad$^{62,48}$, S.~Jaeger$^{4}$, S.~Janchiv$^{25}$, Q.~Ji$^{1}$, Q.~P.~Ji$^{15}$, X.~B.~Ji$^{1,53}$, X.~L.~Ji$^{1,48}$, Y.~Y.~Ji$^{40}$, H.~B.~Jiang$^{40}$, X.~S.~Jiang$^{1,48,53}$, J.~B.~Jiao$^{40}$, Z.~Jiao$^{17}$, S.~Jin$^{34}$, Y.~Jin$^{56}$, M.~Q.~Jing$^{1,53}$, T.~Johansson$^{66}$, N.~Kalantar-Nayestanaki$^{54}$, X.~S.~Kang$^{32}$, R.~Kappert$^{54}$, M.~Kavatsyuk$^{54}$, B.~C.~Ke$^{42,1}$, I.~K.~Keshk$^{4}$, A.~Khoukaz$^{59}$, P. ~Kiese$^{27}$, R.~Kiuchi$^{1}$, R.~Kliemt$^{11}$, L.~Koch$^{29}$, O.~B.~Kolcu$^{52a,m}$, B.~Kopf$^{4}$, M.~Kuemmel$^{4}$, M.~Kuessner$^{4}$, A.~Kupsc$^{66}$, M.~ G.~Kurth$^{1,53}$, W.~K\"uhn$^{29}$, J.~J.~Lane$^{57}$, J.~S.~Lange$^{29}$, P. ~Larin$^{14}$, A.~Lavania$^{20}$, L.~Lavezzi$^{65a,65c}$, Z.~H.~Lei$^{62,48}$, H.~Leithoff$^{27}$, M.~Lellmann$^{27}$, T.~Lenz$^{27}$, C.~Li$^{38}$, C.~H.~Li$^{31}$, Cheng~Li$^{62,48}$, D.~M.~Li$^{70}$, F.~Li$^{1,48}$, G.~Li$^{1}$, H.~Li$^{62,48}$, H.~Li$^{42}$, H.~B.~Li$^{1,53}$, H.~J.~Li$^{15}$, H.~N.~Li$^{46,j}$, J.~L.~Li$^{40}$, J.~Q.~Li$^{4}$, J.~S.~Li$^{49}$, Ke~Li$^{1}$, L.~K.~Li$^{1}$, Lei~Li$^{3}$, P.~R.~Li$^{30,k,l}$, S.~Y.~Li$^{51}$, W.~D.~Li$^{1,53}$, W.~G.~Li$^{1}$, X.~H.~Li$^{62,48}$, X.~L.~Li$^{40}$, Xiaoyu~Li$^{1,53}$, Z.~Y.~Li$^{49}$, H.~Liang$^{62,48}$, H.~Liang$^{1,53}$, H.~~Liang$^{26}$, Y.~F.~Liang$^{44}$, Y.~T.~Liang$^{24}$, G.~R.~Liao$^{12}$, L.~Z.~Liao$^{1,53}$, J.~Libby$^{20}$, C.~X.~Lin$^{49}$, T.~Lin$^{1}$, B.~J.~Liu$^{1}$, C.~X.~Liu$^{1}$, D.~~Liu$^{14,62}$, F.~H.~Liu$^{43}$, Fang~Liu$^{1}$, Feng~Liu$^{6}$, G.~M.~Liu$^{46,j}$, H.~M.~Liu$^{1,53}$, Huanhuan~Liu$^{1}$, Huihui~Liu$^{16}$, J.~B.~Liu$^{62,48}$, J.~L.~Liu$^{63}$, J.~Y.~Liu$^{1,53}$, K.~Liu$^{1}$, K.~Y.~Liu$^{32}$, L.~Liu$^{62,48}$, M.~H.~Liu$^{9,f}$, P.~L.~Liu$^{1}$, Q.~Liu$^{67}$, Q.~Liu$^{53}$, S.~B.~Liu$^{62,48}$, T.~Liu$^{1,53}$, W.~M.~Liu$^{62,48}$, X.~Liu$^{30,k,l}$, Y.~Liu$^{30,k,l}$, Y.~B.~Liu$^{35}$, Z.~A.~Liu$^{1,48,53}$, Z.~Q.~Liu$^{40}$, X.~C.~Lou$^{1,48,53}$, F.~X.~Lu$^{49}$, H.~J.~Lu$^{17}$, J.~D.~Lu$^{1,53}$, J.~G.~Lu$^{1,48}$, X.~L.~Lu$^{1}$, Y.~Lu$^{1}$, Y.~P.~Lu$^{1,48}$, C.~L.~Luo$^{33}$, M.~X.~Luo$^{69}$, P.~W.~Luo$^{49}$, T.~Luo$^{9,f}$, X.~L.~Luo$^{1,48}$, X.~R.~Lyu$^{53}$, F.~C.~Ma$^{32}$, H.~L.~Ma$^{1}$, L.~L. ~Ma$^{40}$, M.~M.~Ma$^{1,53}$, Q.~M.~Ma$^{1}$, R.~Q.~Ma$^{1,53}$, R.~T.~Ma$^{53}$, X.~X.~Ma$^{1,53}$, X.~Y.~Ma$^{1,48}$, F.~E.~Maas$^{14}$, M.~Maggiora$^{65a,65c}$, S.~Maldaner$^{4}$, S.~Malde$^{60}$, Q.~A.~Malik$^{64}$, A.~Mangoni$^{22b}$, Y.~J.~Mao$^{37,h}$, Z.~P.~Mao$^{1}$, S.~Marcello$^{65a,65c}$, Z.~X.~Meng$^{56}$, J.~G.~Messchendorp$^{54}$, G.~Mezzadri$^{23a}$, T.~J.~Min$^{34}$, R.~E.~Mitchell$^{21}$, X.~H.~Mo$^{1,48,53}$, N.~Yu.~Muchnoi$^{10,b}$, H.~Muramatsu$^{58}$, S.~Nakhoul$^{11,d}$, Y.~Nefedov$^{28}$, F.~Nerling$^{11,d}$, I.~B.~Nikolaev$^{10,b}$, Z.~Ning$^{1,48}$, S.~Nisar$^{8,g}$, Q.~Ouyang$^{1,48,53}$, S.~Pacetti$^{22b,22c}$, X.~Pan$^{9,f}$, Y.~Pan$^{57}$, A.~Pathak$^{1}$, A.~~Pathak$^{26}$, P.~Patteri$^{22a}$, M.~Pelizaeus$^{4}$, H.~P.~Peng$^{62,48}$, K.~Peters$^{11,d}$, J.~Pettersson$^{66}$, J.~L.~Ping$^{33}$, R.~G.~Ping$^{1,53}$, S.~Pogodin$^{28}$, R.~Poling$^{58}$, V.~Prasad$^{62,48}$, H.~Qi$^{62,48}$, H.~R.~Qi$^{51}$, M.~Qi$^{34}$, T.~Y.~Qi$^{9}$, S.~Qian$^{1,48}$, W.~B.~Qian$^{53}$, Z.~Qian$^{49}$, C.~F.~Qiao$^{53}$, J.~J.~Qin$^{63}$, L.~Q.~Qin$^{12}$, X.~P.~Qin$^{9}$, X.~S.~Qin$^{40}$, Z.~H.~Qin$^{1,48}$, J.~F.~Qiu$^{1}$, S.~Q.~Qu$^{35}$, K.~H.~Rashid$^{64}$, K.~Ravindran$^{20}$, C.~F.~Redmer$^{27}$, A.~Rivetti$^{65c}$, V.~Rodin$^{54}$, M.~Rolo$^{65c}$, G.~Rong$^{1,53}$, Ch.~Rosner$^{14}$, M.~Rump$^{59}$, H.~S.~Sang$^{62}$, A.~Sarantsev$^{28,c}$, Y.~Schelhaas$^{27}$, C.~Schnier$^{4}$, K.~Schoenning$^{66}$, M.~Scodeggio$^{23a,23b}$, W.~Shan$^{18}$, X.~Y.~Shan$^{62,48}$, J.~F.~Shangguan$^{45}$, M.~Shao$^{62,48}$, C.~P.~Shen$^{9}$, H.~F.~Shen$^{1,53}$, X.~Y.~Shen$^{1,53}$, H.~C.~Shi$^{62,48}$, R.~S.~Shi$^{1,53}$, X.~Shi$^{1,48}$, X.~D~Shi$^{62,48}$, J.~J.~Song$^{40}$, J.~J.~Song$^{15}$, W.~M.~Song$^{26,1}$, Y.~X.~Song$^{37,h}$, S.~Sosio$^{65a,65c}$, S.~Spataro$^{65a,65c}$, K.~X.~Su$^{67}$, P.~P.~Su$^{45}$, F.~F. ~Sui$^{40}$, G.~X.~Sun$^{1}$, H.~K.~Sun$^{1}$, J.~F.~Sun$^{15}$, L.~Sun$^{67}$, S.~S.~Sun$^{1,53}$, T.~Sun$^{1,53}$, W.~Y.~Sun$^{26}$, X~Sun$^{19,i}$, Y.~J.~Sun$^{62,48}$, Y.~Z.~Sun$^{1}$, Z.~T.~Sun$^{1}$, Y.~H.~Tan$^{67}$, Y.~X.~Tan$^{62,48}$, C.~J.~Tang$^{44}$, G.~Y.~Tang$^{1}$, J.~Tang$^{49}$, J.~X.~Teng$^{62,48}$, V.~Thoren$^{66}$, W.~H.~Tian$^{42}$, Y.~T.~Tian$^{24}$, I.~Uman$^{52b}$, B.~Wang$^{1}$, C.~W.~Wang$^{34}$, D.~Y.~Wang$^{37,h}$, H.~J.~Wang$^{30,k,l}$, H.~P.~Wang$^{1,53}$, K.~Wang$^{1,48}$, L.~L.~Wang$^{1}$, M.~Wang$^{40}$, M.~Z.~Wang$^{37,h}$, Meng~Wang$^{1,53}$, S.~Wang$^{9,f}$, W.~Wang$^{49}$, W.~H.~Wang$^{67}$, W.~P.~Wang\href{https://orcid.org/0000-0001-8479-8563}{\includegraphics[scale=0.004]{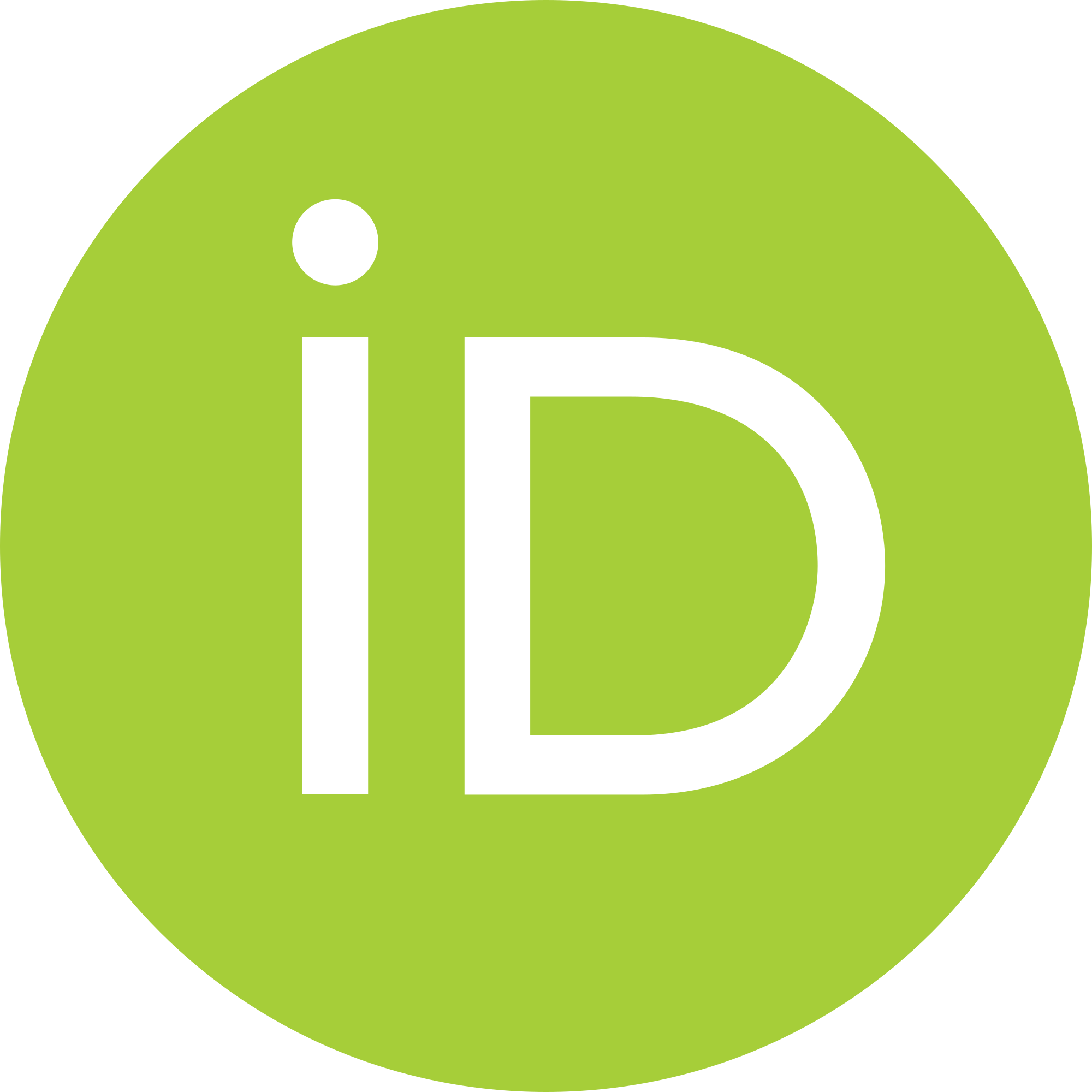}}$^{62,48}$, X.~Wang$^{37,h}$, X.~F.~Wang$^{30,k,l}$, X.~L.~Wang$^{9,f}$, Y.~Wang$^{49}$, Y.~D.~Wang$^{36}$, Y.~F.~Wang$^{1,48,53}$, Y.~Q.~Wang$^{1}$, Y.~Y.~Wang$^{30,k,l}$, Z.~Wang$^{1,48}$, Z.~Y.~Wang$^{1}$
% ...
\\
\vspace{0.2cm}
(BESIII Collaboration)\\
\vspace{0.2cm} {\it
$^{1}$ Institute of High Energy Physics, Beijing 100049, People's Republic of China\\
% ...
$^{9}$ Fudan University, Shanghai 200443, People's Republic of China\\
% ...
$^{30}$ Lanzhou University, Lanzhou 730000, People's Republic of China\\
% ...
$^{36}$ North China Electric Power University, Beijing 102206, People's Republic of China\\
$^{37}$ Peking University, Beijing 100871, People's Republic of China\\
% ...
$^{48}$ State Key Laboratory of Particle Detection and Electronics, Beijing 100049, Hefei 230026, People's Republic of China\\
$^{49}$ Sun Yat-Sen University, Guangzhou 510275, People's Republic of China\\
% ...
$^{53}$ University of Chinese Academy of Sciences, Beijing 100049, People's Republic of China\\
% ...
$^{62}$ University of Science and Technology of China, Hefei 230026, People's Republic of China\\
% ...
$^{f}$ Also at Key Laboratory of Nuclear Physics and Ion-Beam Application (MOE) and Institute of Modern Physics, Fudan University, Shanghai 200443, People's Republic of China.\\
% ...
$^{h}$ Also at State Key Laboratory of Nuclear Physics and Technology, Peking University, Beijing 100871, People's Republic of China.\\
% ...
$^{k}$ Also at Frontiers Science Center for Rare Isotopes, Lanzhou University, Lanzhou 730000, People's Republic of China.\\
$^{l}$ Also at Lanzhou Center for Theoretical Physics, Lanzhou University, Lanzhou 730000, People's Republic of China.\\
% ...
}\end{center}
\vspace{0.4cm}
\vspace{0.4cm}
\end{small}
}
\noaffiliation{}

%% file: Reference.tex

%% file: draft.bbl
\begin{thebibliography}{99}
\bibitem{egconexc}
   For example, the process $\ee\to K^{\ast0}K^{-}\pi^{+}$, which is fully simulated in the {\sc conexc} component of the {\sc hybrid} generator, can also contribute to the implemented inclusive process $\ee\to K^{+}K^{-}\pi^{+}\pi^{-}$ due to the decay $K^{\ast0}\to K^{+}\pi^{-}$. Therefore, this contribution should be subtracted from the inclusive process in advance.
\end{thebibliography}
